\documentclass[times]{rncauth}

\usepackage{moreverb}
\usepackage{cite}
\usepackage[dvips,colorlinks,bookmarksopen,bookmarksnumbered,citecolor=red,urlcolor=red]{hyperref}

\newcommand\BibTeX{{\rmfamily B\kern-.05em \textsc{i\kern-.025em b}\kern-.08em
T\kern-.1667em\lower.7ex\hbox{E}\kern-.125emX}}

\def\volumeyear{2013}

\newcommand{\rtn}{\mathbb{R}}
\newtheorem{remark}{Remark}
\begin{document}

\runningheads{X.~Wang, et al.}{Multi-Agent Distributed Coordination Control: Developments and Directions}

\title{Multi-Agent Distributed Coordination Control: Developments and Directions}

\author{Xiangke  Wang\affil{1}\corrauth, Xun Li\affil{1}, Yirui Cong\affil{1}\affil{2}, Zhiwen zeng\affil{1} and Zhiqiang Zheng\affil{1}\affil{2}}

\address{\affilnum{1}College of Mechantronic Engineering and Automation, National University of Defense Technology, 410073, Changsha, China\break
\affilnum{2} Research School of Engineering, the Australian National University, Canberra ACT 2601, Australia}
\corraddr{College of Mechantronic Engineering and Automation, National University of Defense Technology, 410073, Changsha, China. E-mail: xkwang@nudt.edu.cn.}

\cgsn{the Research Projection of National University of Defense Technology}{JC13-03-02}

\begin{abstract}
In this paper, the recent developments on distributed coordination control, especially the consensus and formation control, are summarized with the graph theory playing a central role, in order to present a cohesive overview of the multi-agent distributed coordination control, together with brief reviews of some closely related issues including rendezvous/alignment, swarming/flocking and containment control.
In terms of the consensus problem, the recent results on consensus for the agents with different dynamics from first-order, second-order to high-order linear and nonlinear dynamics, under different communication conditions, such as cases with/without switching communication topology and varying time-delays, are reviewed, in which the algebraic graph theory is very useful in the protocol designs, stability proofs and converging analysis.
In terms of the formation control problem, after reviewing the results of the algebraic graph theory employed in the formation control, we mainly pay attention to the developments of the rigid and persistent graphs. With the notions of rigidity and persistence, the formation transformation, splitting and reconstruction can be completed, and consequently the range-based formation control laws are designed with the least required information in order to maintain a formation rigid/persistent.
Afterwards, the recent results on rendezvous/alignment, swarming/flocking and containment control, which are very closely related to consensus and formation control, are briefly introduced, in order to present an integrated view of the graph theory used in the coordination control problem.
Finally, towards the practical applications, some directions possibly deserving investigation in coordination control are raised as well.
\end{abstract}


\maketitle
%

\vspace{-6pt}

\section{Introduction}
\vspace{-2pt}
Researchers have long noticed and carried on detailed analysis on many coordinated behaviors, for example, the forage for food or defense against predators of insects, birds and fishes in nature, and self-organization or self-excitation of particles in physics~\cite{Hubbard2004fishschool,Toner2005flocks,Janson2005beeswarms,Couzin2005leadershipanimal,Levine2001Selforganization}.
These drove researchers to consider seriously why the creature and particles take initiative to coordinate, and motivated the studies and applications on multi-agent coordination.
Though it appears to be more complicated than single-agent systems, there are indeed many significant advantages in the coordinations of MAS (multi-agent system) over the single-agent system, for example~\cite{WeiRen2007CSMConsensus,Anderson2008RigidGraphControl,ZYLin2008,WRen2008,Bullo2009,Mesbahi2010}:
\begin{itemize}
  \item distributed sensors and actuators, as well as inherent parallelism;
  \item larger redundancy, higher robustness and great fault tolerance;
  \item performing tasks that single-agent system cannot do;
  \item performing usually the tasks that can be finished by single costly agent with lower cost and  higher excellent performance.
\end{itemize}
No doubt that there are more advantages as well.
In addition, with the improvements of the embedded computing and communicating technologies, and the developments of distributed and decentralized methodologies, the distributed coordinations of MASs have become easy to materialize.
With these inspirations, the problem of distributed coordination control of a network of mobile autonomous agents was of interest in control and robotics in the past decade.

\begin{figure}
\centering
\begin{minipage}[t]{0.4\textwidth}
\begin{center}
\includegraphics [width=1.7in,height=1.2in]{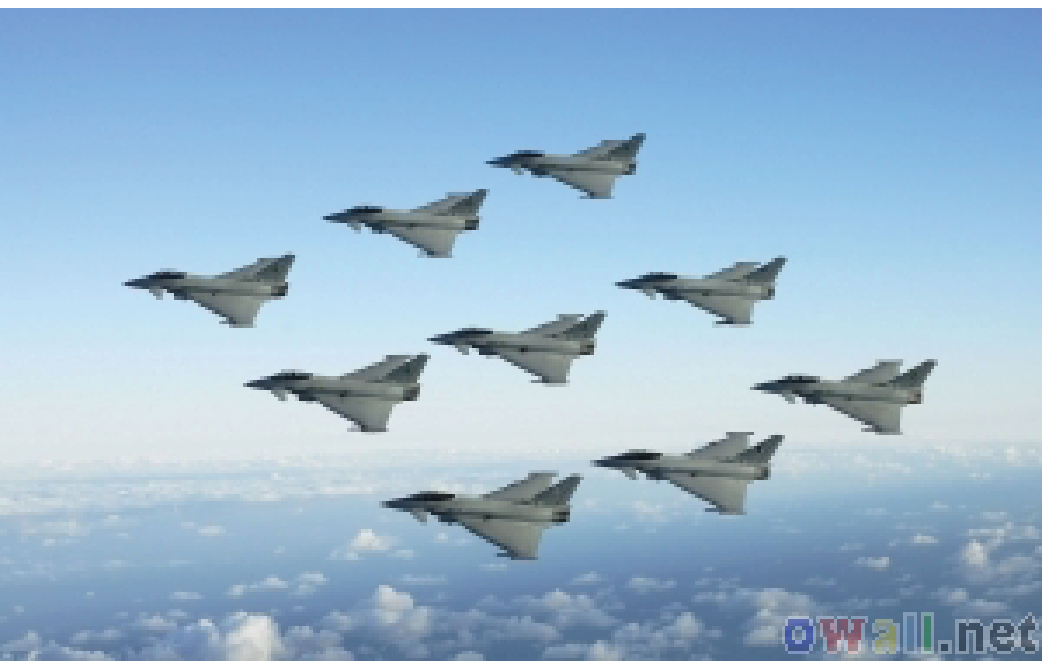}
\end{center}
\end{minipage}
\begin{minipage}[t]{0.4\textwidth}
\begin{center}
\includegraphics [width=1.7in,height=1.2in]{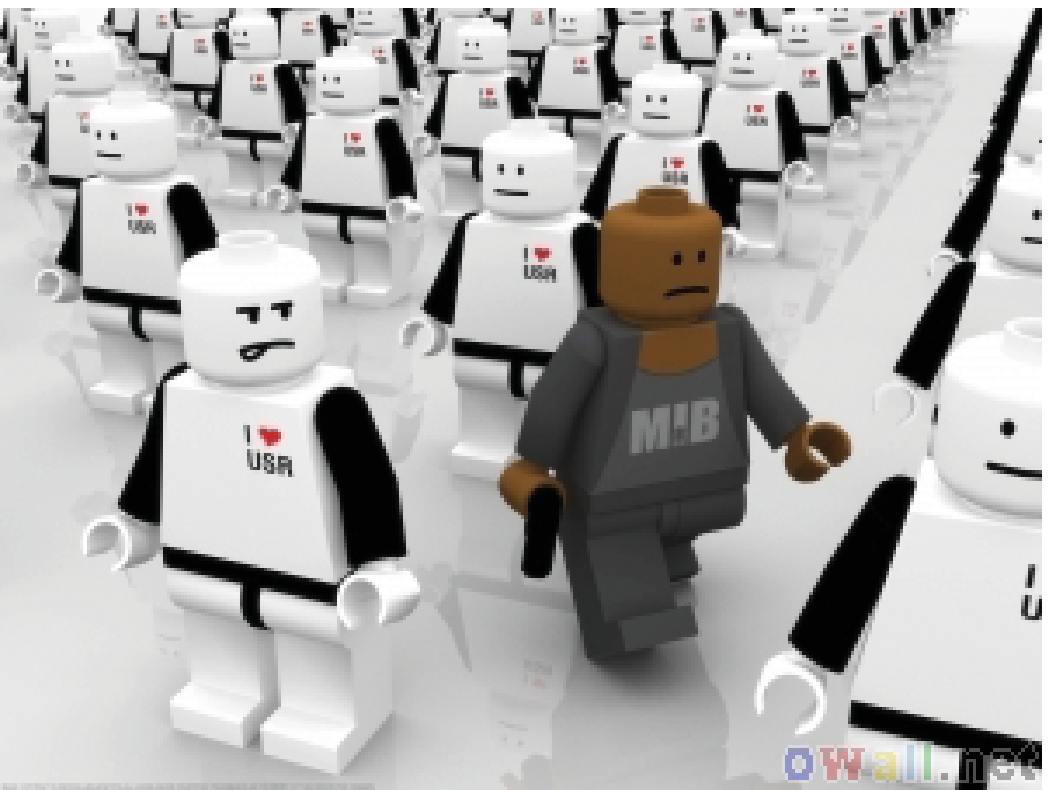}
\end{center}
\end{minipage}\\
\begin{minipage}[t]{0.4\textwidth}
\begin{center}
\includegraphics [width=1.7in,height=1.2in]{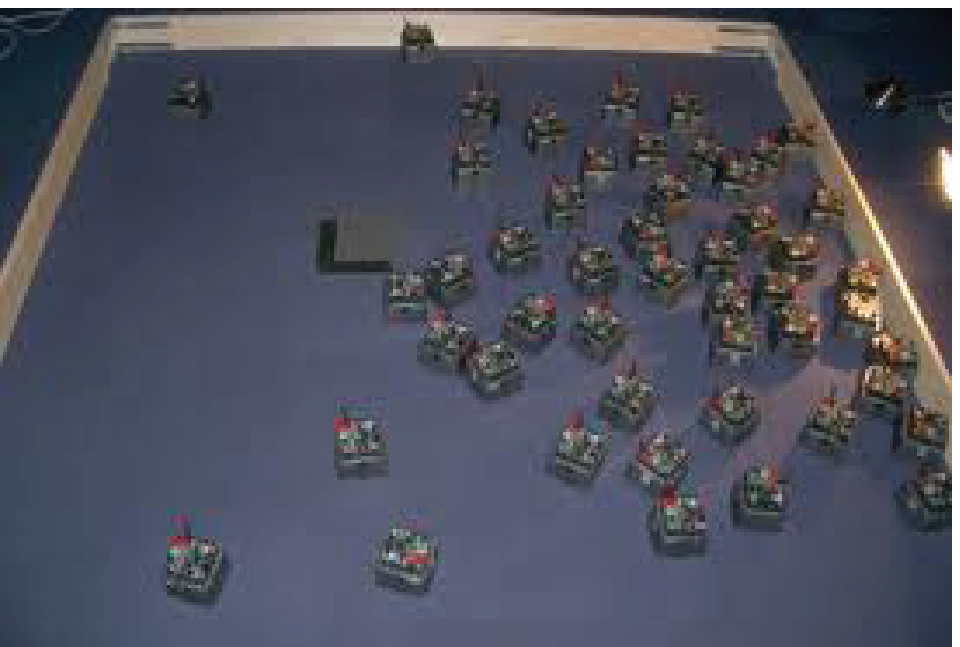}
\end{center}
\end{minipage}
\begin{minipage}[t]{0.4\textwidth}
\begin{center}
\includegraphics [width=1.7in,height=1.2in]{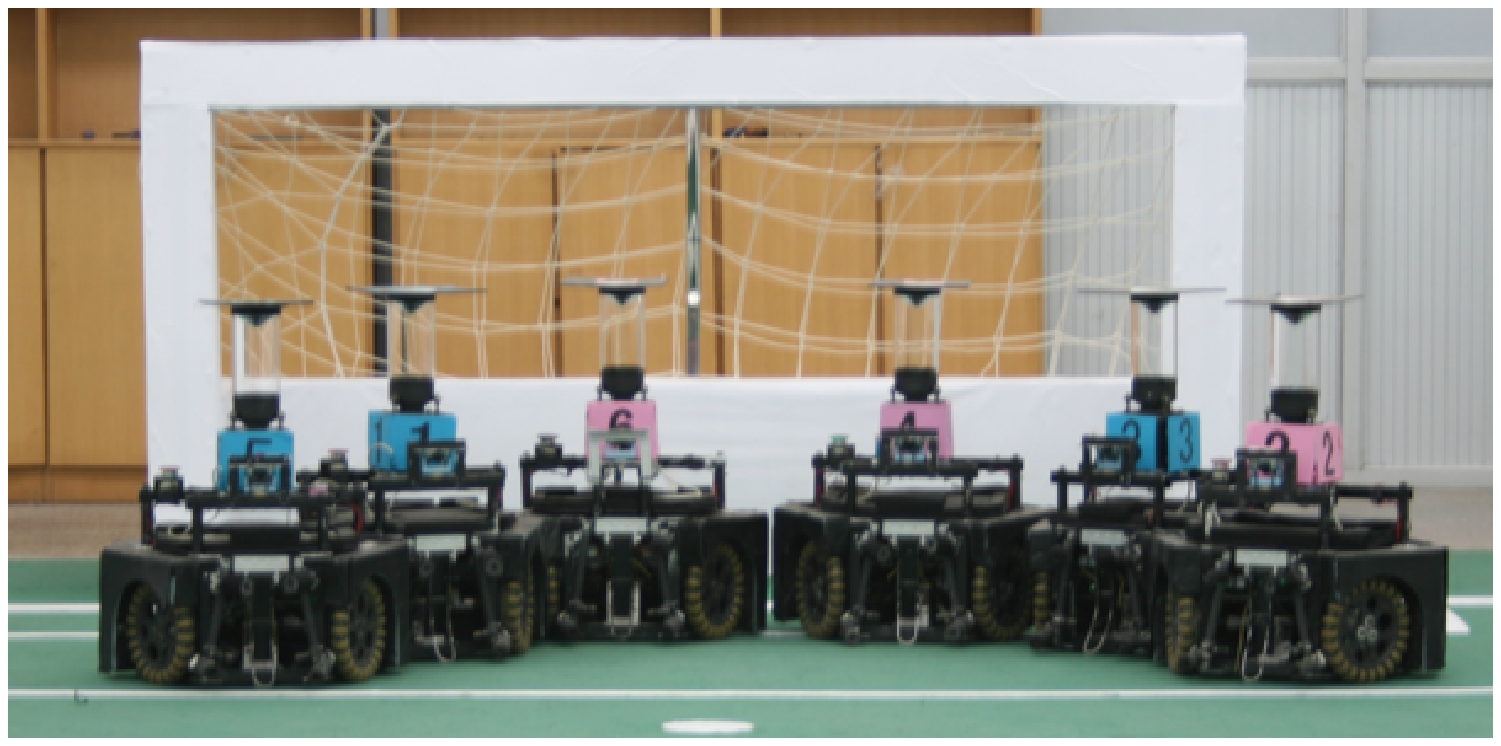}
\end{center}
\end{minipage}
\caption{Some typical applications of multi-agent distributed coordinations}
\label{Fig1}
\end{figure}

Generally speaking, distributed coordination control uses \emph{local} interactions between agents to achieve \emph{collective} behaviors of multiple agents, and therefore to accomplish \emph{global} tasks. It has a broad range of potential applications, such as in military, aerospace, industry, and entertainment. Figure~\ref{Fig1} shows some typical applications of multi-agent distributed coordinations.
In the military field, multiple mobile robot systems can adopt a proper geometric pattern to perform military tasks for taking the place of human soldiers, such as reconnaissance, searching, mine clearance, and patrol under adverse/hazardous circumstances. Taking the reconnaissance mission as an example, a single robot has limited ability to gain environmental information, however, if multiple robots keep proper formation to cooperatively apperceive the surrounding, they are likely to rapidly and accurately obtain the environmental information.
%
In the aerospace field, satellite formation is the leading technique in the space application in 21$^{th}$ century, which opens up a brand-new direction for the application of satellites, especially for mini-satellites. Satellite formation cannot only greatly reduce the cost and enhance the reliability and survivability, but also broaden and override the function of individual satellites and achieve the tasks that multiple single spacecrafts cannot finish.
In the industrial field, multiple mobile robots can deal with some dull, dirty and dangerous work in formation. For example, when multiple robots cooperatively carry large scale goods in a poisonous environment, their positions and orientations are strictly restricted in order to meet the requirements of load balance.
In the entertainment field, for example, multiple dancing robots or soccer robots, in order to keep neat or meet tactical needs, must keep some special patterns, and may dynamically switch the patterns for avoiding obstacles.
%

Distributed coordination control of MASs has attracted tremendous attention with various aspects including consensus, formation,  rendezvous, alignment, synchronization, swarming, flocking, containment control, gossip, cooperative searching and reconnaissance,  cooperative location and mapping,  to name a few.
A number of methods have been proposed for achieving distributed coordination control of MAsS,
and conventional methods include the leader-follower method~\cite{Kumar2001ModelControl,Dasai2002graphmodeling,Luca2008Autinoutconstraints,Dimarogonasa2009cooperativeattitude,Jchen2010IJRRRhS}, behavior-based method~\cite{Balch1998BehaviorbasedFormation,Jadbabaie2003NearestNeighborRules,Antonelli2009,Antonelli2010c}, virtual structure method~\cite{Lewis1997VirtualStructure,Wei2004VirtualStructure,Norman2008,Broek2009FormationVS}, graph-based method~\cite{Murray2004informationflow,Lin2005GraphicalCondition,ChangbinYu2007ThreeandHigherFormation,WeiRen2007CSMConsensus,Anderson2008RigidGraphControl}, etc.
But now, these methods are gradually mixed together, and it is hard to clearly distinguish from each other. In particular, the graph-based method has become the  dominator,  because MASs can be modeled by a graph naturally, and almost all the aspects of coordination control can be then studied by utilizing the graph theory.
Therefore, this paper mainly focuses on the recent developments of consensus, formation control, and some closely related topics such as rendezvous/alignment, swarming/flocking and containment control, with the graph theory paying central roles, in order to present a cohesive overview of the multi-agent distributed coordination control.
Also, some topics which might be interesting for future investigation are discussed.
\vspace{-6pt}
\section{Recent developments\label{sec:developments}}
\vspace{-2pt}
A MAS is in general distinguished with larger scale (often decided by the numbers of agents) and decentralized perception, communication and control structures, and forms inter-connected network between agents consequently.
Thus, it can be naturally modeled by a graph with vertices being used to describe agents and the edges being used to represent topological relationships between agents such as perception, communication and control links.
%
Both directed and undirected graphs can be used to describe the MAS. For example, when the $i^{th}$ agent is required to keep the predetermined distance to the $j^{th}$  agent, while the $j^{th}$  agent is not required to keep the distance to the $i^{th}$  agent,  the edge $\vec{ij}$ is directed from vertex $i$ to $j$, and at this time a directed graph is used to model the MAS. Conversely, when the interconnection between agents $i^{th}$ and $j^{th}$ is bi-directional, when the $i^{th}$ agent could perceive the $j^{th}$ agent if and only if the $j^{th}$ agent could perceive the $i^{th}$ agent for example, the edge $ij$ is undirected, and at this time the graph is also undirected.

\begin{figure}[htbp]
    \begin{center}
        \includegraphics[width=3in]{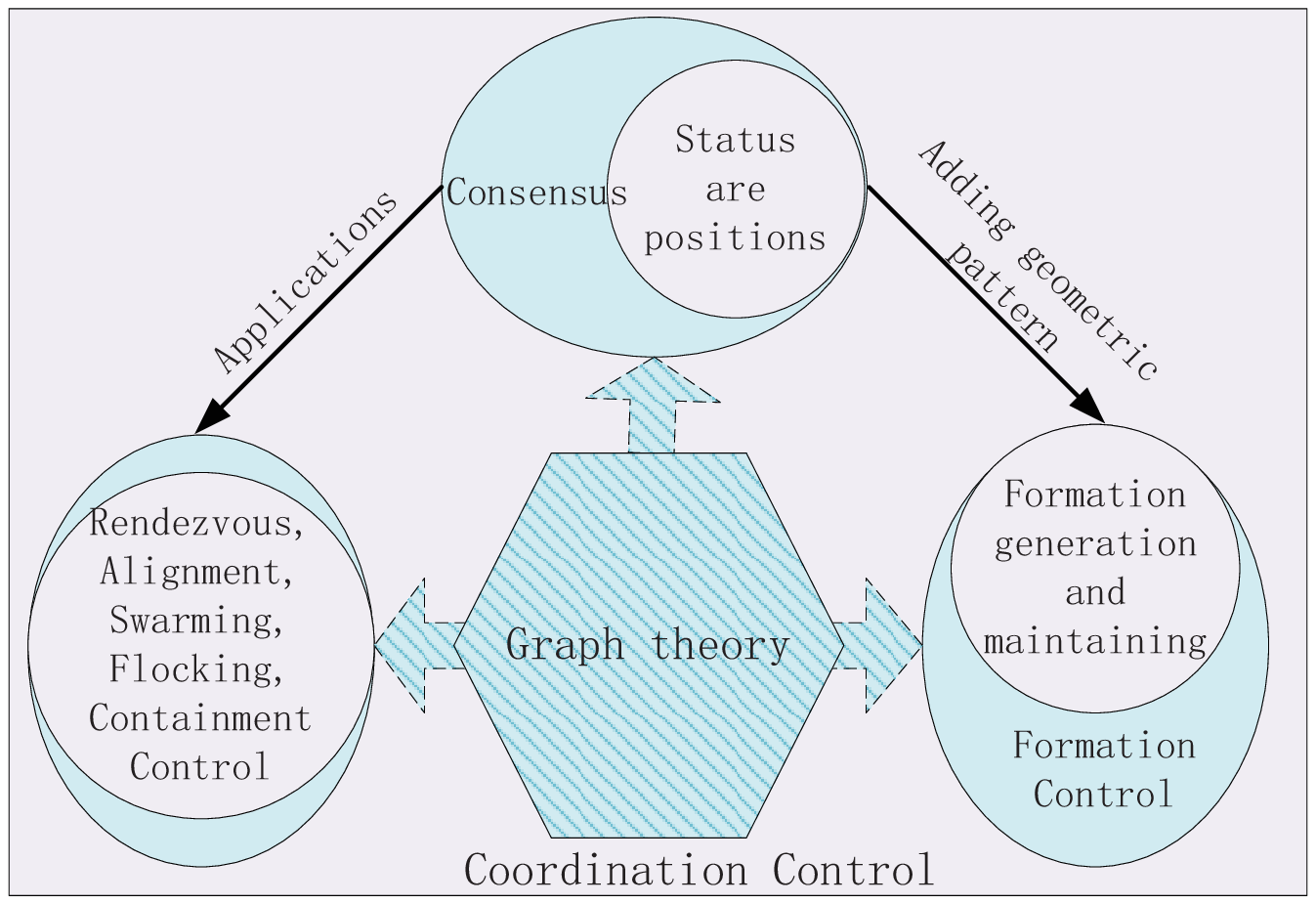}
        \caption{Relationships among consensus, formation control and some closely related topics, i.e.,  rendezvous, alignment, swarming, flocking and containment control.}
        \label{fig:relationships}
    \end{center}
\end{figure}

After modelled by a graph, as noted above, the mature graph theory can be borrowed to study the coordination control of MASs. In recent years, many monographs about the research results of the graph theory on multi-agent coordination control problems, e.g.~\cite{ZYLin2008,WRen2008,Bullo2009,Mesbahi2010}, are published.
Therefore, in the sequel, we will mainly pay attention on the survey of recent developments on some typical multi-agent coordination problems, including  consensus, formation control, rendezvous/alignment, swarming/flocking and containment control, with graph theory playing a central role.
The relationships among the investigated coordination control problems are summarized as Fig.~\ref{fig:relationships}.

\subsection{Consensus}

In the networks of agents, the \emph{consensus} means to reach an agreement regarding a certain
quantity of interest that depends on the states of all agents.
A consensus \emph{protocol} is an interaction
rule that specifies the information exchange between an
agent and all of its neighbors in the network~\cite{Saber2007ConsensusCooperation,WeiR2007CSMConsensus}.

\begin{figure}[htbp]
\begin{center}
\includegraphics [width=4.5in]{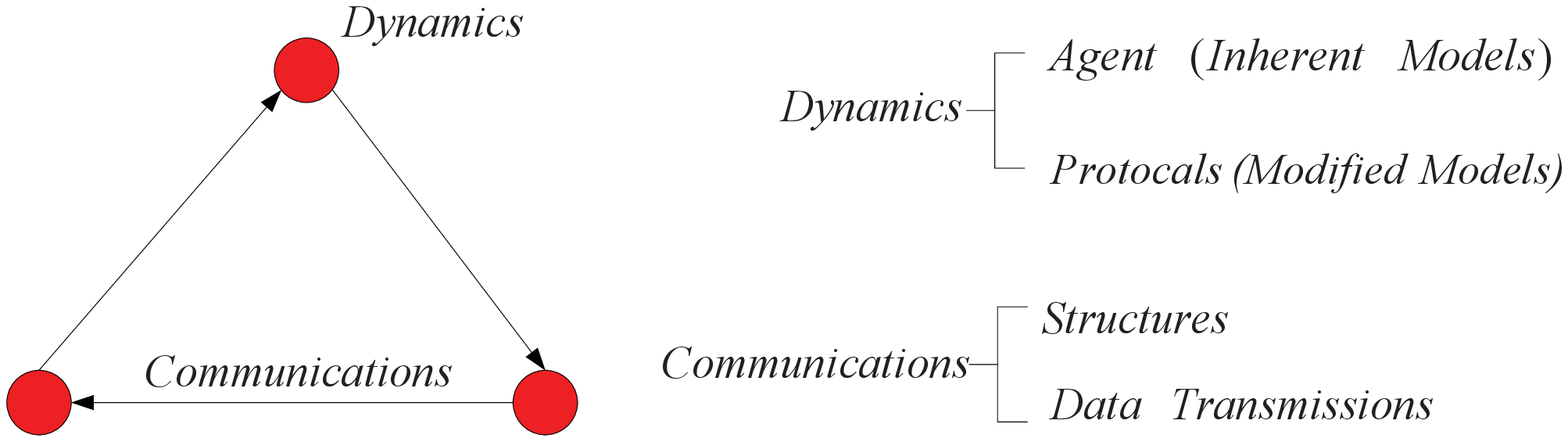}
\caption{Key elements considered in consensus problems}
\label{fig:conscls}
\end{center}
\end{figure}

In general, there are two key elements considered in consensus problems, that is, the dynamics of agents and the communications among agents, as illustrated in Fig.~\ref{fig:conscls}. To be more precise, two parts are normally taken into account in the agents' dynamics, which are the dynamics of the agent itself and the protocol used among agents. The former is the inherent model of the agents, such as the first/second/high-order models, linear/nonlinear models, and continuous/discrete models;  the latter is used to modify the agent inherent model in order to make the MAS achieve a consensus.
The communications among agents can also be divided into communicating structures such as the shapes and changing of the communication topology, and  data transmissions including time delays and transmission failure.


On the other hand, the consensus problems can also be divided into two categories by considering the predominance of agents. If an agent has priority so that the other agents must follow its trajectory, then such an agent is called the ``leader''; accordingly, if the MAS has at least one leader, then the consensus is leader-follower consensus; otherwise, it is leaderless consensus. Even though it seems there is an obvious difference between leader-follower and leaderless consensus, they have the same foundations, because the topology and consensus protocol can make a common agent act as a leader. Afterwards, we will mainly  focus on the developments of leaderless consensus and then point out some important results about leader-follower consensus.


The beginning of researches on consensus is~\cite{Vicsek1995particles}, in which Vicsek et al. propose a simple discrete-time model of particles all moving in the plane at the same speed but with different headings.
Afterwards,  the \emph{algebraic graph theory}, which studies the relationships between the structure of graphs and different matrix representations of graphs, is widely employed in the consensus~\cite{Godsil2004algebraicgraph}.
The most important concepts of the algebraic graph theory used in MASs are the \emph{adjacent matrix} and the \emph{Laplacian matrix}.
For a graph $\cal{G}=(\cal{V},\cal{E})$ consisting of a nonempty
\emph{vertex} set ${\cal{V}}=\{1,\ldots,N\}$ and an \emph{edge} set
$\cal E\subseteq V\times V$, the \emph{adjacency matrix} ${\cal A(G)}=[a_{ij}]$ is an $N\times{N}$
matrix given by $a_{ij}=1$, $a_{ij}>0$ if and only if $(i,j)\in {\cal E}$, and $a_{ij}=0$,
otherwise.
The \emph{degree} of vertex $i$ is given by $d_i=\sum_ja_{ij}$, and the \emph{Laplacian matrix} of $\cal G$
is ${\cal} L_n({\cal G})=\mathrm{diag}(d_1,\ldots,d_N)-{\cal A(G)}$.
The pioneering work to deal with consensus by virtue of the Laplacian matrix is literature~\cite{Jadbabaie2003NearestNeighborRules}, in which  a theoretical explanation for the observed behavior in ~\cite{Vicsek1995particles} is provided.
After that a large number of literatures use the algebraic graph theory to study the consensus problem, for example, literature~\cite{Murray2004TACConsensus,Zlin2004TACLocalControl,WRen2005TACConsensus,WeiR2007CSMConsensus,Daniel2011TACEdgeAgreement,Zlin2007SICONAgreement,MCao2008SICOConsensus,Rahmani2009SICONControllability,YSu2012Switchconsensus,Moreau2005TACcommunication}, to name a few.

The most basic case of consensus is that the considered agents are governed by linear dynamics, especially by single-integrator dynamics.
%
%
The theoretical framework for reaching a consensus for networked agents with single-integrator dynamics under fixed/switching communicating topology was introduced by Olfati-Saber and Murray in literature~\cite{Murray2004TACConsensus}, building on their earlier work of~\cite{Murray2004informationflow}.
The subsequent studies mainly followed up on the way of~\cite{Murray2004TACConsensus}. For single-integrator dynamics, the commonly used continuous-time consensus protocol is ~\cite{Zlin2004TACLocalControl,Murray2004TACConsensus,WRen2005TACConsensus,WeiR2007CSMConsensus,YSu2012Switchconsensus}
\begin{equation}\label{consensuscontinousalgorithm}
\dot{x}_i(t)=-\sum\limits^n_{j=1}a_{ij}(t)\bigl(x_j(t)-x_i(t)\bigr),~i=1,\ldots,n;
\end{equation}
for first-order discrete systems, the commonly used discrete-time consensus protocol is \cite{Jadbabaie2003NearestNeighborRules,WRen2005TACConsensus,Moreau2005TACcommunication,WeiR2007CSMConsensus}
\begin{equation}\label{consensusdiscretealgorithm}
x_i[k+1]=\sum\limits^n_{j=1}a_{ij}[k]x_j[k],~i=1,\ldots,n,
\end{equation}
in which $x_i(t)$ and $x_i[k]$ represent the  states of the $i^{th}$ agent, $a_{ij}(t)$ and $a_{ij}[k]$ are the $(i, j)$ entry of the adjacent matrix of the associated communication graph of MASs at time $t$ or $k(t)$.
Note that protocols~\eqref{consensuscontinousalgorithm} and~\eqref{consensusdiscretealgorithm} only use the relative state information of the adjacent agent, so it is local and distributed.
Then, by using the tools of the algebraic graph theory, different conclusions for different cases are made.
Under the condition of the fixed topology, protocols~\eqref{consensuscontinousalgorithm} and~\eqref{consensusdiscretealgorithm} can guarantee a consensus if the communication graph is connected for an undirected graph or has a rooted spanning tree for a directed graph.
By contrast, when the topology is switching, protocols~\eqref{consensuscontinousalgorithm} and~\eqref{consensusdiscretealgorithm} result in a consensus if there exists an infinite sequence of contiguous, uniformly bounded time intervals,
with the property that across each interval, the \emph{union} of the communication graphs is connected for an undirected case or has a rooted directed spanning tree for a directed case.
Furthermore, by using the maximum and minimum eigenvalues of the Laplacian matrix, the converging rate of the consensus protocol is analyzed~\cite{Lxiao2004lineariteration}.

Time delays associated with both message transmission
and processing, for instance caused by the communication congestion, are inevitable and also taken into account in consensus problems.
Let $\delta_{ij}$ denote the time delay communicated from
the $j^{th}$ agent to the $i^{th}$ agent, protocol~\eqref{consensuscontinousalgorithm}  is modified as
\begin{equation}\label{consensuscontinousdelayalgorithm}
\dot{x}_i(t)=-\sum\limits^n_{j=1}a_{ij}(t)\big(x_j(t-\delta_{ij})-x_i(t-\delta_{ij})\bigr).
\end{equation}
Reference~\cite{Murray2004TACConsensus} investigated the simplest case where $\delta_{ij} = \delta$ and the communication
topology was fixed, undirected, and connected, and concluded that the consensus
was achieved if and only if $0\le\delta<\pi/(2\lambda_{\max}(L))$,
where $L$ denoted the Laplacian of the communication graph. Along the way developed by~\cite{Murray2004TACConsensus},
different time delays and communication topologies are considered in the consensus problem.
If the time delay affects only the state  being transmitted,  protocol~\eqref{consensuscontinousdelayalgorithm}  is modified as
\begin{equation}\label{consensuscontinousdelayalgorithm2}
\dot{x}_i(t)=-\sum\limits^n_{j=1}a_{ij}(t)\bigl(x_j(t-\delta_{ij})-x_i(t)\bigr),
\end{equation}
and the consensus for switching directed topologies remains valid for an arbitrary time delay when $\delta_{ij} = \delta$~\cite{Moreau2004CDC}.
The constant or time-varying, uniformly or non-uniformly distributed time delays are considered in~\cite{Blimana2008consensusdelayed}, and
sufficient conditions for the existence of average consensus under bounded communication
delays are given correspondingly.
The consensus problem with noise-perturbed  under fixed and switching topologies
as well as time-varying communication delays is investigated in~\cite{Sun2011Consensusnoisy}.
It is shown that the consensus is reached for arbitrarily large constant, time-varying,
or distributed delays if consensus is reached without delays~\cite{Munz2011TAC}.
For the discrete consensus protocol~\eqref{consensusdiscretealgorithm}, conclusions similar  to that of the continuous case can be obtained.
It is shown in~\cite{Tanner2005CDC} that if a consensus is reached under a time-invariant undirected communication topology, then the presence of
communication delays does not affect the consensus.
In addition, sufficient
conditions for the consensus under dynamically changing communication topologies and bounded time-varying
communication delays are shown in reference \cite{FXiao2006IJC}.

The results of the first-order system have been extended to the consensus of the second-order and high-order linear dynamical MASs, and similar consensus protocols and graph conditions are obtained.
%
Consensus for multiple agents governed by the second-order linear dynamics has been considered by the similar framework of using the algebra graph theory~\cite{LWang2007ConsensusDynamic,HSu2008AutomaticaSynchronization,JQin2011AutomaticaConsensus,YCAO2012TACVSC}. For example, in the representative work~\cite{LWang2007ConsensusDynamic},  a linear distributed consensus protocol for the second-order MAS is designed with aid of the Laplacian matrix without requiring velocity information of neighbors; a variable structure control law is used to design the consensus protocol by taking the second-order system as two cascade fist-order systems in~\cite{YCAO2012TACVSC}.
In terms of high-order consensus problems, the first discussion is completed by W.~Ren in~\cite{WRen_Moore_2006IEEEInternationalConferenceonNetworking}, where the second-order MAS is generalized to the $l^{th}$-order integrator MAS. Afterwards, the consensus protocol is modified in \cite{FCJiang_GMXie_2008AsianJournalofControl} and $\chi$-consensus is investigated under undirected communication topology.
It should be noted that both the aforementioned literatures have no self-feedback information in the consensus protocols. With adding the self-feedback items, reference~\cite{FCJiang_LWang2010InternationalJournalofControl} investigates the constant-value consensus problem  for high-order MASs under fixed and switching directed topology. Based on those meaningful explorations, the robust analysis and the converging results with time delays are further provided,  for example, the work in~\cite{YLiu_YMJia2010InternationalJournalofRobustandNonlinearControl, LMo_YJia2011IETControlTheoryandApplications, PLin_ZLi_2011IETControlTheoryandApplications}.
%
The consensus governed by more complex high-order linear dynamics, such as the SISO (single input single output) and the MIMO (multiple inputs multiple outputs) linear dynamics are also investigated. At the beginning, researcher did not notice the relationship between the consensus in MAS and the controllability in individual agents; therefore, many studies are done with the assumption that the isolated agent is controllable, e.g. the work in ~\cite{Seo_SHim_2009Automatica}. It is proven that this technical assumption is not necessary in~\cite{JXXi_NCai_2010PhysicaA}, excluding the necessity of
a sufficient connection of the graph topology is required for consensus stability for  a system with unstable agent dynamics. Afterwards, the studies on communication delays and robustness are conducted, e.g. \cite{YLiu_YMJia2012InternationalJournalofSystemsScience, YZhang_YPTian2013IJSS}. And further, a necessary and sufficient condition for high-order consensus with unknown communication
delays is given for the existence of solution to heterogeneous multi-agent systems in~\cite{YPTian2012Automatica}.

Considering that most physical systems are inherently nonlinear in
nature, the consensus where the agents are governed by nonlinear dynamics has also aroused the attention of some researchers recently. It can be shown, for example,
in~\cite{YZhao2013TSMCBHConsensus,HQLi2013TSMCBConsensus} where the
continuously differentiable nonlinear dynamics are considered;
in \cite{WenwuYu2010Secondorderconsensus, QSong2010SCLSecondorderconsensus, QSong2013TSMCBConsensus} where the nonlinear dynamics
satisfy the global Lipschitz condition under a directed communicating graph, respectively, without and with a leader agent; and in
\cite{Abhijit2010IJRNCSychronization} where an adaptive control
method is introduced to study the synchronization of uncertain
nonlinear networked systems.

\subsection{Formation Control}

The \emph{formation control} is that a team comprised of multiple agents keeps a predetermined geometric pattern and adapts to the environmental constraints (e.g. obstacle avoidance) at the same time during the movement towards a specific goal.
In general, the main control problems in formation are summarized as follows~\cite{LM2005,RDH2005}:
\begin{itemize}
\item	Formation generation: how to design a formation pattern for MASs and then achieve it?
\item	Formation maintaining: how to maintain the formation pattern for MASs during the movement?
\item	Formation transformation: how to transit the formation pattern, which means transiting one formation pattern into another?
\item	Obstacle avoidance with formation: how to change a motion plan or a formation pattern for MASs in order to avoid obstacles during movement?
\item	Self-adaptation: how to automatically change the formation in order to best adapt to the dynamical unknown environment?
\end{itemize}

The graph theory is a powerful tool to study formation control  involving all
the above five aspects.
Firstly, graphs are naturally used to generate and maintain a formation,  and consequently to achieve transformation between different formation patterns.
In early 2001, directed graphs were used to describe the topologies  and patterns of formation, and then study the formation evolution problem~\cite{Kumar2001ModelControl}.
Saber et al adopted the combination graph theory to obtain a unique determined formation pattern through designing the weights of edges~\cite{Olfati-Saber02graphrigidity}.
Desai et al in~\cite{Dasai2002graphmodeling} further adopted directed acyclic graphs to represent the control graph between agents, and designed the control strategy, afterwards, through enumerating all possible control graphs to realize transformation between any two formation patterns.
It is worthy pointing out that by adding a pre-specified geometric pattern, the formation generation and maintaining  can also be achieved in the
corresponding consensus problem, which is usually known as \emph{consensus-based formation control}, e.g., in~\cite{WRen2006ACC,WGWL2007ConsensusBasedFormation}.
Consequentially, the algebra graph theory is widely employed in formation control  too.
As aforementioned, the neighbors of  vertex $i$ in the graph represented MAS is exactly the collection of agents that have a topological relationship such as perception with the $i^{th}$ agent.
Therefore, by using the properties of the Laplacian matrix, local, distributed and scalable formation controllers can be designed, and their stabilities can also be verified by virtue of the eigenvalue of the Laplacian matrix, for a typical example, the work in the classic literature~\cite{Lin2005GraphicalCondition}.
And then, Fax and Murray set up the relation between the formation controller and the topological structure of the communicating network with the Laplacian matrix, and proved that if the local controller was stable, then the stability of formation with linear dynamics depended on the stability of the information flow~\cite{Murray2004informationflow}; Lin  et.~al proved that if and only if there is a global accessible vertex in the perception graph, the formation is stable by using tools of the algebraic graph theory~\cite{Lin04localcontrol}.
In addition, the formation stability concepts are defined with the aid of the graph theory.
Tanner et.~al defined the \emph{leader-to-formation stability} notion based on the graph and input-to-state stability concept, which includes both transient and steady-state errors~\cite{Tanner2004LTFStability}.
In contrast, the \emph{pairwise asymptotic stability} is defined based on directed graphs by involving a single pair of neighboring agents in~\cite{XKWANG2012TRO},
which implies that any two agents can asymptotically achieve and maintain a desired relative attitude and position though the pair of agents which may not be neighbors and do not interact with each other directly.

\begin{figure}[htbp]
    \begin{center}
        \includegraphics[width=5in]{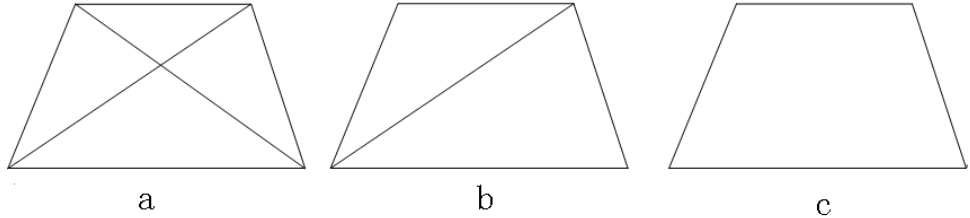}
        \caption{(a) rigid graph ~~(b) minimally rigid graph ~~(c) non-rigid graph}
        \label{fig:rigidgraph}
    \end{center}
\end{figure}

\begin{figure}[htbp]
\begin{center}
\includegraphics [scale=1.0,trim=0 0 0  0]{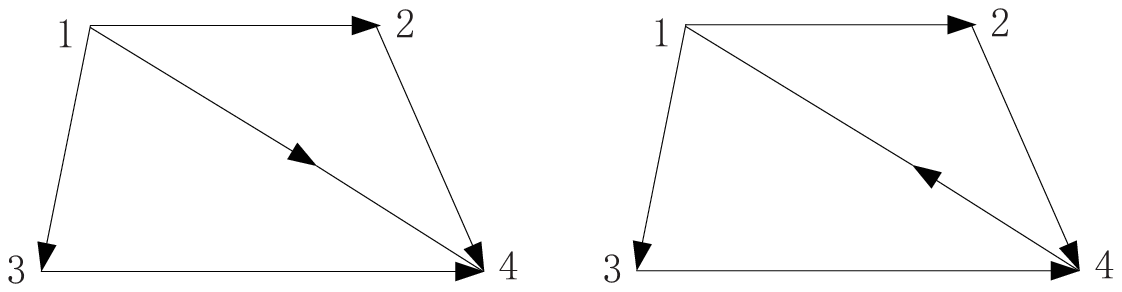}
\caption{(a) nonpersistent graph in $\rtn^2$~~~~~~(b) persistent graph in $\rtn^2$}
\label{Fig3}
\end{center}
\end{figure}

Another important tool of the graph theory employed in multi-agent formation control is the rigid graph theory.
An undirected graph is \emph{rigid} if for almost every structure, the only possible continuous moves are those which preserve every inter-agent distance;
further, an undirected graph is called \emph{minimally rigid} if it is rigid and
if there exists no rigid graphs with the same number of vertices
and a smaller number of edges (see Fig.~\ref{fig:rigidgraph} for more illustrations of \emph{rigid} and \emph{minimally rigid} graphs).
%
%
The main tools of rigid graphs are Henneberg sequence and Laman's theorems~\cite{Graver1994CombinatorialRigidity}. The former was raised by Henneberg to construct two-dimensional minimally rigid graphs, and the latter were raised by Laman in 1970 to verify if a two-dimensional graph was rigid~\cite{Laman1970rigidity}.
With the concepts of rigid graphs, some primitive operators are defined to deal with the operations such as splitting and restructuring of rigid formation~\cite{Olfati-Saber02graphrigidity,Eren2004Operations}; also, a formation control law is proposed by only using the distances between agents in order to ensure rigidity or generic rigidity of formation~\cite{Murray2004informationflow}.
Note that the concept of rigidity is mainly for the undirected graph, which means the interrelations between vertices are bi-directional.
However, in practical applications, the MAS is often modeled by a directed graph for reducing the cost of communication and perception.  Hence on the basis of rigidity, a group led by Prof. Brian D.O. Anderson at the Australian National University developed a ``directed rigidity'' concept, named  \emph{persistence}.
A directed graph is \emph{persistent} if and only if its underlying graph is rigid, and itself is constraint consistent ~\cite{ChangbinYu2007ThreeandHigherFormation}, as illustrated in Fig.~\ref{Fig3}~\footnote{
In $\rtn^2$, the graph represented in (a) is not persistent. For
almost all uncoordinated displacements of 2, 3 and 4 (even if they satisfy their
constraints), 4 is indeed unable to satisfy its three constraints. This problem
cannot happen for the graph represented in (b), which is persistent however.}. Similar to a minimally rigid one, a graph is \emph{minimally persistent} if it is persistent and if no single edge can be removed without losing persistence.
The main difference between rigidity and persistence is that rigidity assumes all the constraints are satisfied, as if they were enforced
by an external agency or through some mechanical properties, while persistence considers each
constraint to be the responsibility of a single agent.
Based on the concept of persistence,  many works have been investigated by the group led by Prof. Brian D.O. Anderson, such as developing the persistent/minimum persistent notions from the rigid/minimum rigid concepts~\cite{ChangbinYu2007ThreeandHigherFormation}; fixing the transformation, splitting and reconstruction of formation in a two-dimensional space through defining three primitive graph operators by virtue of Henneberg sequence~\cite{Hendrickx2008,Hendrickx2007}; designing a formation control law to keep the persistence for formation in two-dimensional space~\cite{ChangbinYu2009PersistentFormations}; and further proposing distributed formation control laws in order to keep the minimum persistence for two kinds (leader-remote-follower and coleader) of formation with small perturbations~\cite{Summers2011TACPersistent}.
It is worthy pointing out that the persistence notions have been applied to practical multiple robots systems. For example, B.~Smith et al designed an automatic generation algorithm of persistent formations for multiple robots under distance and communication constraints~\cite{Brian2007}, and a method for accomplishing such persistent formations was presented and further demonstrated with a prototype network of robots in a NASA project for researches in Antarctica~\cite{Brian20082}.

\subsection{Some closely related issues}

\subsubsection{Rendezvous/Alignment}

The \emph{rendezvous} problem refers to designing a distributed local control strategy to make multiple agents reach the same specified position at the same time. To some extent, rendezvous can be taken as the application of consensus in actual systems, such as robots and spacecrafts~\cite{Kranakis2006rendezvoussurvey}.
This problem was firstly proposed in literature~\cite{Ando1999distributedmemorylessconvergence}, in which a distributed simple memoryless point rendezvous algorithm was proposed with proven convergence.
And then, literature~\cite{Lin2007rendezvoussynchronous} and~\cite{Lin2007rendezvousasynchronous} extended this algorithm to a
``stop-and-go'' strategy under synchronous and asynchronous situations, respectively, and analyzed the convergence of the strategies.
Similar to other multi-agent coordinations, the graph theory also played key roles in the research of the rendezvous problem.
Cortes et.~al presented a robust rendezvous algorithm in an arbitrarily dimensional space with the aid of the proximity graph~\cite{Cortes2006rendezvousproximitygragh};
and a connectivity-preserving protocol consisting of a set of distributed control rules was proposed in the case of the link in communication graph failure in literature~\cite{FXiao2012Automatica}.
In addition, in view of different kinds of practical systems, the rendezvous problem has received extensive researches.
For multiple omni-directional mobile robots equipped with line-of-sight limited-range sensors moving in a connected, nonconvex,
unknown environment, literature~\cite{Ganguli2009rendezvouswithvisibilitysensors} presented a  perimeter minimizing
rendezvous algorithm by using the notions of connectivity-preserving constraint sets and proximity graphs.
For multiple nonholonomic unicycles, a discontinuous and time-invariant rendezvous algorithm was designed with  tools from the nonsmooth Lyapunov theory and the graph theory~\cite{Dimarogonas2007redezvousnonholonomic}.
The discontinuous rendezvous policies were further investigated in literature~\cite{Conte2010redezvousdiscontinuouscontrol}, in which
the new proposed sufficient conditions for characterizing the control policies were less restrictive than
those  presented in the above-mentioned literature.
For a multiple agent system moving like a Dubins car, a simple
quantized control law was designed with three values to make agents achieve rendezvous given a connected initial assignment with minimalism in sensing and control~\cite{JYu2012TACRendezvous}.

The \emph{attitude alignment} problem, sometimes called attitude consensus or attitude synchronization problem, also received extensive attention in multi-agent fields, especially multiple spacecrafts~\cite{Ren2007formationkeepingatitudealignment}, multiple satellites ~\cite{Bondhus2005synchronizationsatelliteattitude} and multiple marine robots~\cite{Smith2001orientationcontrol}.
Similar to the rendezvous problem, the attitude alignment is required to design a distributed control strategy to make the attitude of multiple agents tend to be consistent simultaneously. Lawton et al~\cite{Lawton2002synchronizedultiplespacftrotations} adopted the behavior-based method to design two kinds of control strategies that made a group of aircrafts achieve attitude alignment; and further, this work was extended to more general scenes which did not need bi-directional communication~\cite{Ren2007formationkeepingatitudealignment}. For the attitude alignment with limited communication and no reference signal, Sarlette et al proposed attitude alignment strategies based on artificial potential methods~\cite{Sarlette2009attitudesynchronization}.

\subsubsection{Swarming/Flocking}

The \emph{swarming} is often inspired by biological and physical systems, which refers to the prevalent collective behavior and the self-organization phenomenon, such as ant colonies, bee colonies, flocks of birds and schools of fishes~\cite{Couzin2005leadershipanimal,Toner2005flocks,Janson2005beeswarms}. And recently it has emerged in MAS with focuses on physical embodiment and realistic interactions among the individuals themselves and also between the individuals and the environment~\cite{Barca2012swarmreviewed}.
%
%
Generally speaking, swarming is characterized by the following: 1) flexibility: adaptability to the environment; 2) robustness: anti-jamming to the internal and external disturbances; 3) dispersion: dynamical behavior based on individuals; 4) self organization: obvious overall system properties in evolution, namely emerging~\cite{DZChen2004Swarming}.

As a special case of swarming, the \emph{flocking} refers to the phenomena that the MAS (usually with the second-order integrator dynamics) presents certain coordinated behaviors by decentralized control with the aid of local perception  and reaction between individuals.
The classical flocking model is proposed by Reynolds in~\cite{Reynolds1987}, in which the animation of flocking behaviors, such as bird flock and fish school, are realized through three rules, i.e. collision avoidance, velocity matching and flock centering.
These three rules can make the distance between agents converge to an expected value, the speed of agents tend to be consistent, and agents cannot collide with each other.
After then, for the linear systems with the second-order integrator dynamics, flocking algorithms are usually designed by local artificial potential fields integrated with the graph theory~\cite{Saber2006Flocking,SWL2009Flocking,Tanner2007Flocking}.
In~\cite{Saber2006Flocking}, a theoretical framework based on the algebraic graph
theory and the spatially induced graphs is provided for the design and analysis of scalable flocking
algorithms under a connected topology.
Further, this work was extended in \cite{SWL2009Flocking} from two directions: one is the case where only a fraction of agents are informed and the other is that where the  velocity of the virtual leader is varying.
In \cite{Tanner2007Flocking}, the stability properties of a group of agents governed by decentralized, nearest-neighbor interaction rules are analyzed.
The flocking problem without considering collision between multi-agents has also got extensive attentions in recent years, e.g. in~\cite{Lee2007TACFlocking,Moshtagh2007TACFlocking}.
While little work has been done on flocking with nonlinear dynamics, except that
Dong proposed a backsteping based control law design
method for the flocking of multiple nonholonomic wheeled mobile robots with directed topology in~\cite{Dong2011Flocking}.

\subsubsection{Containment control}

The \emph{containment control} can be taken as an expansion of consensus or flocking with pinning control, which
refer to designing a distributed control law to drive the states of the followers to the convex hull spanned by the states of multiple leaders. In particular, when the multiple leaders are moving, some literature call it as \emph{set tracking} problem~\cite{GRSHI2012TACSettracking}.
In early 2005, the simplest containment control, which took two agents as leaders and drove the rest converge to the line decided by the two leaders, was discussed in ~\cite{Lin2005GraphicalCondition}.
For the containment control with the first-order dynamics under fixed undirected graphs,  a simple ``stop-and-go'' strategy is proposed and analyzed in  literature~\cite{MJi2008TACcontainmentcontrol}.
Further, the containment control with single- or double-integrator dynamics was studied in~\cite{MJi2008TACcontainmentcontrol,YCao2011TCSTcontainmentcontrol,Ycao2009CDC,JLi2012TACcontainmentcontrol,YLU2012Automatica},
of which, literature~\cite{JLi2012TACcontainmentcontrol} proposed two containment control algorithms
via only position measurements where the leaders were neighbors of only a subset of the followers,
and literature~\cite{YLU2012Automatica} examined the containment control  for a second-order multi-agent system with random switching topologies.
Moreover, the containment control strategy for multiple Lagrangian system in directed topological structure was proposed in~\cite{JMei2012Automaticacontainmentcontrol}; and the containment control of multiple rigid-body systems with uncertainty was studied in ~\cite{ZMeng2010Automaticaattutudecontainmentcontrol}.
For the more general nonlinear dynamics, G. Shi et al~\cite{GRSHI2012TACSettracking} investigated the distributed set tracking problem with
unmeasurable velocities under switching directed topologies, and provided the necessary and sufficient conditions for set input-to-state stability and set integral input-to-state stability.

\begin{remark}
The studies on coordination control are still ongoing, both in depth and width. On the one hand, the studied agents are found with more and more complicated dynamics and limited perceptions, for example, from high-order dynamics~\cite{YPTian2012Automatica} to complex environmental constraints~\cite{Ganguli2009TRONonconvex}, again to communication bandwidth limitation~\cite{TLi2011limitedbandwidth,TLi2011TAC,Javad2012QuantizedConsensus}, and quantitative communication~\cite{Kashyap2007QuantizedConsensus}.
On the other hand, more and more attention has been paid to various practical multi-robot systems, for example, multiple micro-satellites~\cite{Rune2011SpacecraftSynchronization}, multiple spacecrafts~\cite{Michael2011transportation,Muller2011Juggling}, multiple marine robots~\cite{Dudek2011IROSboat}, multiple wheeled mobile robots~\cite{WJDONG2012TRO,Luca2012HierarchicalFormations}, and multiple Lagrangian systems~\cite{Silvia2012Automatica}.
\end{remark}
\vspace{-6pt}
\section{Directions}
\vspace{-2pt}
The theoretical work on multi-agent coordination control has made significant achievements in the past decade, which have however rare practical applications.
In fact, the behaviors of actual robots are much more complicated than those considered in the existing work.
Therefore, how to promote the theoretical work for serving in practice in order to shrink the gap between theory and practice is still an open problem which should be considered seriously.
In the authors' opinion, towards this direction, there are several topics deserving further investigation in the study of multi-agent coordination control, which listed as follows:

1) Coordination control with strongly nonlinear dynamics

The existing coordination control often considers the agents governed by linear dynamics, especially by the single- or double-integrator
dynamics; however, most physical systems are inherently nonlinear in nature.
At present, coordination control with  nonlinear dynamics has received seldom consideration, and few related work assume the nonlinear dynamics to be continuously differentiable \cite{YZhao2013TSMCBHConsensus,HQLi2013TSMCBConsensus} or globally Lipschitz~\cite{WenwuYu2010Secondorderconsensus,QSong2010SCLSecondorderconsensus}.
In contrast, the dynamics of robots, vehicles or UAVs in practice are generally found with strong nonlinearities, which are hard to be described simply by using single/double-integrator or continuously differentiable/global Lipschitz continuous function.
So it is necessary and beneficial to study multi-agent coordination
control in the presence of the nonlinearities by means of the nonlinear control theory. Literature~\cite{YGHong2010CDC,XKWANG2012IETCTA} carried on some preliminary research on this way  via the input-to-state stability concept, but more problems, such as the finite time convergence, converging rate analysis, with topological switching and time-delay, are still worthy of further exploration.

2)	Coordination control in three-dimensional space

The existing work about multi-agent coordinations usually considers agents in a plane; while the coordinations in the more general three-dimensional space, composed of three rotational and three translational degrees of freedoms,  are not extensively investigated.
In essence, any general rigid motion (including rotation and translation) is described in three-dimensional space and its subspaces,
and the coordination control in a plane is just a special case in three-dimensional space.
Moreover, the applications of multi-agent coordination control are not confined to be in the plane. For instance, in the applications of multiple spacecraft formation flights~\cite{Rune2011SpacecraftSynchronization}, multiple marine robots cooperative exploration~\cite{Dudek2011IROSboat}, flight array~\cite{Andrea2011flightarray}, and multiple flying vehicles cooperative handling objects~\cite{Michael2011transportation} or playing tennis~\cite{Muller2011Juggling}, the coordination controls are required to be considered in the three-dimensional space with attitude and position control simultaneously.
 Such problems are worthwhile to be solved, in the authors' opinion, for coordination control in three-dimensional space include two aspects:
a) the graph theory in three-dimensional space. As noted above, the graph theory is an important tool used in multi-agent coordination control, however, the existing results on the graph theory cannot be directly extended to the three-dimensional space. For example, Laman's theorem and Henneberg sequence in three-dimensional space are still open problems~\cite{ChangbinYu2009PersistentFormations}. Improving and expanding the existing foundations of the graph theory in order to make them valid in three-dimensional space is still a problem that has not been effectively solved.
b) designing the coordination control law in a three-dimensional space. In recent years, different mathematical tools have been used to study the multi-agent coordination control in the three-dimensional space, such as the decoupling method (namely carrying on the attitude control and the position control independently)~\cite{Muller2011Juggling,Rune2011SpacecraftSynchronization}, the Lie-group abstraction~\cite{Michael2009PlanningControl}, the homogeneous transformation matrix~\cite{Igarashi2009SynchronizationSE(3),Takeshi2012PoseSynchronization}, and the dual quaternion approach~\cite{XKWANG2012TRO}, etc. Whatever, the above methods are still not systematic,  and it is necessary to combine with the graph theory to further design coordination control laws under different topology structures, with different dynamic models and adaptation to external disturbances.

3) Coordination control with practical constraints

In practice, there are many constraints for the coordinations of MASs, such as non-ideal communications and perceptions, and heterogeneous dynamics. Challenges from pure theories to practical applications should be attached with more extensive concern, in order to shrink the gap between theories and applications.

Communications are very important for multi-agent coordinations. Communications for actual systems are usually not ideal.For example, time-delays, unstable signals, and limited bandwidth are inevitable in the actual systems. Therefore, multi-agent coordination control with non-ideal communication is a challenging problem, and many work can be further considered, such as the design of fault tolerance topological structure which is robustness against the loss of communicating nodes or links, coordination algorithms with limited communicating bandwidth or quantitative communication, or even without communications, and underlying communication protocols with reliable robustness.

In addition, it is usually the ideal perceptions that is considered by the existing MASs, which means agents can obtain all the required information in real-time. However in practice, sensors equipped by agents certainly have some perceptional limitations. For example, the commonly used cameras have conical perceptional fields, and the laser ranging sensor (typically the Hokuyo's series products) can only perceive the information in a sector.
Meanwhile, the perceived information is usually accompanied by noise and time-delay, and further more information, such as other agents' relative velocities and accelerations, is generally difficult to obtain precisely. Therefore, the coordination control with perceptional constraints is also a problem that requires to be fixed, such as coordination control with perceptional directions, coordination control without velocity measurement, and coordination control with time-delay and noisy information.

Thirdly, in the existing work on coordination control,  the agents are usually considered to be isomorphic, which means all agents are found with the same dynamics. However, in many practical applications, for example, in the cooperative reconnaissance carried out by manned/unmanned aerial vehicles, and the cluster spacecraft system, the dynamics between agents are quite different.
So, the coordination control for heterogeneous MASs is also a direction worth to pay attentions to in the future.

Finally,  most of the existing theoretical results are only verified by simulations, rather than by actual systems, partly due to the high cost and various restrictions of the experiments. Therefore, to verify and apply the theoretical results to actual multi-robot systems is a most pressing issue too.

4) Combination with other collective behaviours

For the last ten years, the work on coordination control in fact has led the research of the distributed networked system and provided some necessary supports to different types of collective behaviors, for e.g. the wireless sensor network.  Therefore, how to expand and fuse the existing results of coordination control to other collective behaviours such as wireless sensor network is also a notable direction.

\vspace{-6pt}
\section{Conclusion}\label{summary}
\vspace{-2pt}
This paper reviews the recent developments on multi-agent coordination control with focus on the consensus, formation control, and some closely related topics including rendezvous/alignment, swarming/flocking and containment control,  with the graph theory playing a central role,  in order to present a cohesive overview of the multi-agent distributed coordination control, and further provides some directions possibly deserving to be investigated in coordination control.
The work on coordination control involves extremely extensive contents, and the authors believe that in the coming decade, more problems related to coordination control will be solved along with the developments of other related disciplines and technologies, and will also get more practical applications in more fields.

\bibliographystyle{wileyj}

\bibliography{survey}

%
%
%
\end{document}